\documentclass{aa}  

\usepackage{graphicx}
\usepackage{txfonts}
\usepackage{xcolor}
\usepackage{url}
\usepackage{rotating}
\usepackage{amsmath,amssymb,CJK}

\providecommand{\myeol}{\\}
\providecommand{\okina}{`}
\providecommand{\oum}{1I/{{\okina}Oumuamua}}
\providecommand{\borisov}{2I/Borisov}

\providecommand{\vinf}{\ensuremath{v_\infty}}
\providecommand{\solused}{47}
\providecommand{\water}{H$_2$O}

\providecommand{\gaia}{Gaia}
\providecommand{\gdr}[1]{Gaia~DR{#1}}
\providecommand{\gmag}{\ensuremath{G}}
\providecommand{\mg}{M$_\gmag$}
\providecommand{\bprp}{BP-RP}
\providecommand{\teff}{\ensuremath{T_{\rm eff}}}

\providecommand{\tenc}{\ensuremath{t_{\rm enc}}}
\providecommand{\denc}{\ensuremath{d_{\rm enc}}}
\providecommand{\venc}{\ensuremath{v_{\rm enc}}}

\providecommand{\tencmed}{\ensuremath{t_{\rm enc}^{\rm med}}}
\providecommand{\dencmed}{\ensuremath{d_{\rm enc}^{\rm med}}}
\providecommand{\vencmed}{\ensuremath{v_{\rm enc}^{\rm med}}}

\providecommand{\ra}{\ensuremath{\alpha}}
\providecommand{\dec}{\ensuremath{\delta}}

\providecommand{\parallax}{\ensuremath{\varpi}}

\providecommand{\propm}{\ensuremath{\mu}}
\providecommand{\vx}{\ensuremath{v_x}}
\providecommand{\vy}{\ensuremath{v_y}}
\providecommand{\vz}{\ensuremath{v_z}}
\providecommand{\sigmavx}{\ensuremath{\sigma(\vx)}}
\providecommand{\sigmavy}{\ensuremath{\sigma(\vy)}}
\providecommand{\sigmavz}{\ensuremath{\sigma(\vz)}}
\providecommand{\corvxvy}{\ensuremath{\rho(\vx, \vy)}}
\providecommand{\corvxvz}{\ensuremath{\rho(\vx, \vz)}}
\providecommand{\corvyvz}{\ensuremath{\rho(\vy, \vz)}}
\providecommand{\vr}{\ensuremath{v_r}}

\providecommand{\ms}{\ensuremath{\textrm{m\,s}^{-1}}}
\providecommand{\kms}{\ensuremath{\textrm{km\,s}^{-1}}}
\providecommand{\maspyr}{\ensuremath{\textrm{mas\,yr}^{-1}}}
\providecommand{\msun}{\ensuremath{M_\odot}}
\providecommand{\mjup}{\ensuremath{M_{\rm Jup}}}
\providecommand{\degree}{\ensuremath{^\circ}}

\begin{document}
\begin{CJK*}{UTF8}{gbsn}

\title{A search for the origin of the interstellar comet 2I/Borisov}
\titlerunning{Origins of 2I/Borisov}
\authorrunning{Bailer-Jones et al.}
\author{Coryn A.L.\ Bailer-Jones\inst{1}\thanks{calj@mpia.de}, Davide Farnocchia\inst{2}, Quanzhi Ye (叶泉志)\inst{3},
Karen J. Meech\inst{4}, Marco Micheli\inst{5,6}}
\institute{Max Planck Institute for Astronomy, K\"onigstuhl 17, 69117 Heidelberg, Germany \and
Jet Propulsion Laboratory, California Institute of Technology, 4800 Oak Grove Drive, Pasadena, CA 91109, USA \and
Department of Astronomy, University of Maryland, College Park, MD 20742 \and
Institute for Astronomy, 2680 Woodlawn Drive, Honolulu, HI 96822 USA \and
ESA NEO Coordination Centre, Largo Galileo Galilei 1, 00044 Frascati (RM), Italy \and
INAF - Osservatorio Astronomico di Roma, Via Frascati 33, 00040 Monte Porzio Catone (RM), Italy
\\
}
\date{Submitted 2 December 2019; Resubmitted 13 and 27 December 2019; Accepted 30 December 2019}
\abstract{
  The discovery of the second interstellar object \borisov\ on 2019 August 30 raises the question of whether it was ejected recently from a nearby stellar system. Here we compute the asymptotic incoming trajectory of \borisov, based on both recent and pre-discovery data extending back to December 2018, using a range of force models that account for cometary outgassing. From \gdr{2} astrometry and radial velocities, we trace back in time the Galactic orbits of 7.4 million stars to look for close encounters with \borisov. The closest encounter we find took place 910\,kyr ago with the M0V star \object{Ross 573}, at a separation of 0.068\,pc (90\% confidence interval of 0.053--0.091\,pc) with a relative velocity of 23\,\kms. This encounter is nine times closer than the closest past encounter identified for the first interstellar object \oum.  Ejection of \borisov\ via a three-body encounter in a binary or planetary system is possible, although such a large ejection velocity is unlikely to be obtained and \object{Ross 573} shows no signs of binarity. We also identify and discuss some other recent close encounters, recognizing that if \borisov\ is more than about 10\,Myr old, our search would be unlikely to find its parent system.  }
 \maketitle

\section{Introduction}\label{sec:intro}

The discovery of the first interstellar object, \oum, on 2017 October 19 generated an intense period of observation using over 100 hours of medium to large telescope time for the two weeks following its discovery. The strong interest in \oum\ resulted from this being the first opportunity to acquire detailed information on a fragment of material ejected from another star system.
The scientific community was eager for the discovery of a second interstellar object because it would start to provide information on the diversity of interstellar objects.

The data obtained on \oum\ immediately following its discovery focused on the characterization of its physical properties, while the final data, obtained by the Very Large Telescope and Hubble Space Telescope between November 2017 and early January 2018, was an astrometric experiment designed to trace the orbit back to \oum's parent solar system. From these observations we learned that \oum\ was small and red, which is typical of solar system comets, but looked asteroidal \citep[without dust or gas,][]{meech2017, ye2017, fitzsimmons2018}. The extreme brightness variations showed that the nucleus was rotating in an excited state and was highly elongated -- something that is yet to be explained \citep{meech2017,belton2018}. Because of the lack of activity, it was surprising when the astrometry showed a strong non-gravitational signal that was perturbing the motion of \oum\ as it left the solar system, suggesting that there was undetected cometary outgassing \citep{micheli2018}.  The non-Keplerian trajectory was compared with the reconstructed orbits of 7.4 million stars from the Gaia Data Release 2 (\gdr{2}) catalog to try to identify past close encounters that might be \oum's parent star system, but no low velocity close encounters were found \citep{bailer-jones2018}.

The enduring interest in interstellar objects has stimulated a rich field of investigation, with more than 120 papers written on \oum\ to date.  Yet because of the brevity of its visit, we still lack information regarding some of the most fundamental questions about \oum\ regarding its composition, shape, and origin.

Comet \object{C/2019 Q4 (Borisov)} was discovered at low elevation in the morning twilight on 2019 August 30 by Gennady Borisov at the MARGO observatory in Crimea \citep{borisov2019} when it was at 2.98\,au from the Sun and exhibited a short dust tail.  After analyzing the astrometry reported to the Minor Planet Center (MPC) in the following week, the possibility of a highly hyperbolic orbit became apparent. Because it was moving more slowly than \oum, it took longer for the orbit to be recognized as interstellar. However, by 2019 September 10 our astrometric data from the Canada-France-Hawaii telescope showed a hyperbolic orbit at the 10-$\sigma$ level with an eccentricity of 3.08 (later refined to 3.36).  The hyperbolic orbit was published by the MPC on 2019 September 11 \citep{mpcR106} and this was officially designated as \borisov\ on 2019 September 24 \citep{mpcS71}. \borisov\ is red \citep{2019ApJ...886L..29J} with similar colors to \oum; our data give $g-r$\,=\,$0.64\pm 0.01$\,mag.  Unlike \oum, \borisov\ is actively outgassing, and CN has been detected, giving it a chemical composition that appears to be similar to solar system carbon-chain species depleted comets \citep{fitzsimmons2019,Opitom2019}.

Pre-discovery observations subsequently reported by the Zwicky Transient Facility and going back as far as 2018 December 13 \citep{ye2019}, brought the total observed arc to almost a year (we use here observations up to 2019 November
17) with a total of 916 observations.\footnote{\url{https://www.minorplanetcenter.net/db_search/show_object?utf8=&object_id=2I}}
In this paper we use these data to compute the asymptotic trajectory of \borisov, and use astrometry from the \gdr{2} catalogue \citep{2018A&A...616A...1G} plus radial velocities to trace stellar positions back in time, in order to identify any close encounters.
We cannot perform a survey that is ``complete'' in any useful sense of that term, because we are limited by the available \gaia\ astrometry and, more significantly, by the available matching radial velocities (see details below).
Thus our goal is limited to a search for close encounters among specific stars. We do not attempt
to correct for incompleteness in a statistical manner.

\section{Asymptotic trajectory}\label{sec:trajectory}

To determine \borisov 's asymptotic incoming trajectory we fitted a suite of different models (listed below) to account for the different possible behavior of non-gravitational forces.
\begin{itemize}
\item Gravity-only (JPL solution 38).
\item \cite{marsden1973} model, in which the radial, transverse, and out-of-plane components of the non-gravitational perturbation are $A_i g(r)$ for $i=1,2,3$ respectively. The $A_i$ parameters are constant while the function $g(r)$, which describes the dependency on the heliocentric distance $r$, is driven here by \water. JPL solution 39 estimates only $A_1$ and $A_2$, solution 40 also estimates $A_3$, and solution 41
estimates all of these as well as a time offset relative to perihelion of the peak of the nongravitational acceleration \citep{1989AJ.....98.1083Y}.
\item \cite{marsden1973} model with $g(r)$ driven by CO \citep{2004come.book..317M}. JPL solution 42 estimates only $A_1$ and $A_2$, whereas solution 43 also estimates $A_3$.
\item Rotating jet model \citep{1950ApJ...111..375W,1951ApJ...113..464W,chesley2005} with $g(r)$ driven by \water. The two jets are located at 150\degree\ and 135\degree\ of colatitude.
There are two distinct minima for the pole: JPL solution 44 corresponds to a pole at (\ra, \dec) = (150\degree, $-$55\degree), solution 45 to (\ra, \dec) = (315\degree, 25\degree).
\item Rotating jet model, with $g(r)$ driven by CO. JPL solution 46 corresponds to a pole at (\ra, \dec) = (205\degree, $-$55\degree), solution 47 to (\ra, \dec) = (340\degree, 30\degree) \citep{ye2019}. 
\end{itemize}

\begin{table*}
\begin{center}
  \caption{Solutions for the incoming asymptotic trajectory (velocity vector) of \borisov. (\ra, \dec) and \vinf\ give the barycentric ICRF (International Celestial Reference Frame) direction and asymptotic velocity of \borisov.  \sigmavx, \sigmavy, \sigmavz\ are the standard deviations of the Cartesian ICRF velocity components; the last three columns give the correlation coefficients $\rho$.
For orientation, the Galactic coordinates of these solutions are $l=132.9\degree, b=-1.9\degree$.  We use only solution \solused\ in the analysis.
\label{tab:asymptote_solutions}
}
\begin{tabular}{cccccccccc}
\hline
Solution & \ra   & \dec  & \vinf  & \sigmavx & \sigmavy & \sigmavz & \corvxvy & \corvxvz & \corvyvz \\
      & [deg] & [deg] & [\kms] & [\kms]   & [\kms]   & [\kms]   &          &          &          \\
\hline
38 &    32.79806 &    59.44090 &   32.284055 &   0.0001812 &    0.0002747 &    0.0004150 &       0.8210 &       0.8735 &       0.9878 \\
39 &    32.79823 &    59.44022 &   32.284287 &   0.0003626 &    0.0003430 &    0.0005600 &       0.8465 &       0.8937 &       0.9761 \\
40 &    32.79809 &    59.44016 &   32.284313 &   0.0003774 &    0.0003438 &    0.0005601 &       0.7925 &       0.8632 &       0.9725 \\
41 &    32.79751 &    59.44008 &   32.283096 &   0.0005672 &    0.0005446 &    0.0011744 &       0.8654 &       0.9090 &       0.9664 \\
42 &    32.80272 &    59.44094 &   32.279114 &   0.0006349 &    0.0003488 &    0.0011396 &       0.3297 &       0.8514 &       0.6946 \\
43 &    32.79926 &    59.43894 &   32.285134 &   0.0017506 &    0.0006598 &    0.0022294 &       0.8543 &       0.9588 &       0.9173 \\
44 &    32.79747 &    59.44015 &   32.282620 &   0.0005058 &    0.0005272 &    0.0010957 &       0.8971 &       0.9430 &       0.9636 \\
45 &    32.79747 &    59.44015 &   32.282644 &   0.0004957 &    0.0005208 &    0.0010804 &       0.8989 &       0.9437 &       0.9646 \\
46 &    32.79798 &    59.44021 &   32.284561 &   0.0018604 &    0.0015284 &    0.0041306 &       0.9886 &       0.9912 &       0.9925 \\
47 &    32.79720 &    59.44014 &   32.286894 &   0.0024122 &    0.0011258 &    0.0041225 &       0.9282 &       0.9886 &       0.9500 \\
  \hline
\end{tabular}
\end{center}
\end{table*}

\begin{figure}
\begin{center}
\includegraphics[width=0.5\textwidth, angle=0]{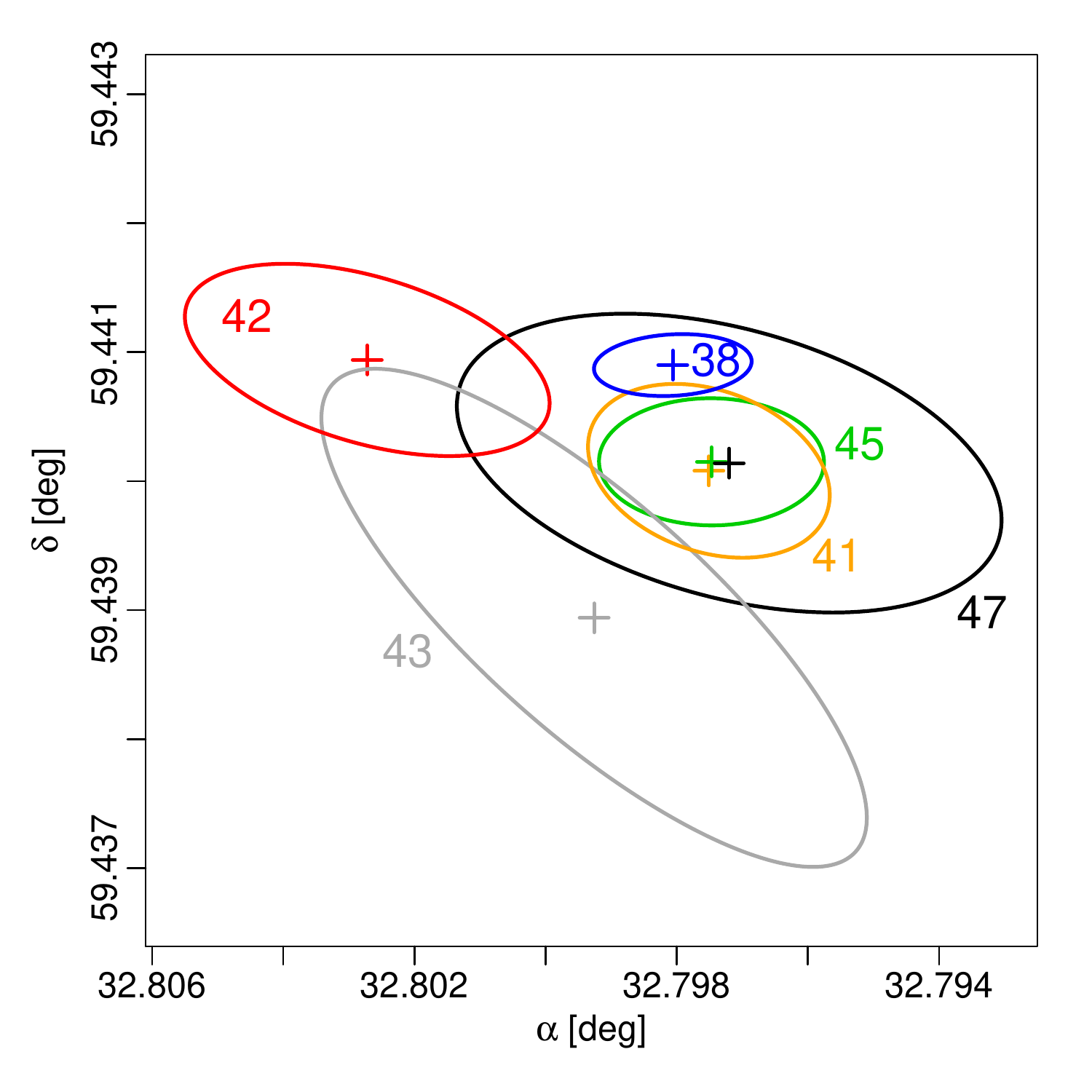}
\caption{Right ascension (\ra) and declination (\dec) marked by a cross, plus uncertainty ellipse (90\% confidence region), of the incoming asymptotic trajectory of \borisov. The labels refer to the JPL solution numbers in Table~\ref{tab:asymptote_solutions}. Both axes span the same angular range of $\Delta\ra\cos(\dec)=\Delta\dec=0.00691\degree$.
\label{fig:asymptote}
}
\end{center}
\end{figure}

Similarly to what \cite{bailer-jones2018} did for \oum, for each orbital solution we numerically integrated the trajectory back to 3000 BCE and then extrapolated to the Keplerian inbound asymptote relative to the solar system barycenter.
The gravitational model takes into account the Sun, the eight planets, the Moon, Pluto, 16 main belt perturbers, and general relativity \citep{2015aste.book..815F}.
The parameters for these asymptotic solutions (giving the direction \borisov\ was coming from, and its speed, before encountering the Sun), including complete uncertainties, are listed in Table~\ref{tab:asymptote_solutions}. A selection is plotted in Figure~\ref{fig:asymptote}.

For the rest of our analysis we use only solution \solused. The covariance for this solution in Table~\ref{tab:asymptote_solutions}  corresponds to 1$\sigma$ uncertainties in \ra, \dec, and \vinf\ of 3.5\arcsec, 1.4\arcsec, and 0.0049\,\kms\ respectively.
This solution captures reasonably well the overall scatter caused by the different non-gravitational models. The maximum difference in the direction of the different asymptotic trajectories is 12\arcsec, which corresponds to a transverse deviation of 0.002\,pc over a path of 30\,pc (the distance \borisov\ travelled to the closest encounter we find below). The maximum difference in the asymptotes' velocity magnitudes is 8\,\ms, corresponding to a displacement of 0.008\,pc along the path after 1\,Myr. Both of these are small compared to the uncertainties in the stellar trajectories, as we see from the results.
While revising this paper, further JPL solutions became available, yet these are all consistent with JPL solution \solused\ to well within the uncertainties, differing by less than 1\,\ms\ and at most a few arcseconds. No further significant change of the incoming asymptote is expected.

\section{Close stellar encounters}

To identify close encounters, we integrate the orbits of a sample of stars back in time through a Galactic potential, and we do the same for \borisov\ starting from its asymptotic trajectory computed in the previous section. Our sample comprises all sources in \gdr{2} that have five-parameter astrometry, and radial velocities either from \gdr{2} or (if not in \gdr{2}) from Simbad \citep{2000A&AS..143....9W}.
This is mostly limited to cool stars (roughly 3500--7000\,K) with $5<\gmag<14$\,mag.
We further limit our search to objects that have {\tt visibility\_periods\_used}\footnote{This is the number of distinct observation epochs, defined as an observation group separated from other groups by a gap of at least four days.}
\,$ \geq 8$ and unit weight error (UWE) less than 35 (see section 2.2 of \citealt{2018A&A...616A..37B} for further explanation). In total our  sample contains 7\,428\,838 sources.\footnote{This total consists of three groups: 7\,039\,430 sources have all six parameters from \gdr{2}; 337\,767 sources have no radial velocity in \gdr{2} but do have a radial velocity and a \gdr{2} {\tt source\_id} in Simbad (i.e.\ they were cross matched by CDS); 51\,641 sources have a radial velocity in Simbad but no \gdr{2} {\tt source\_id} in Simbad, yet nonetheless have a close ($<1\arcsec$) counterpart in \gdr{2}. This last group includes some duplicates of sources in the first group, but uses a Simbad as opposed to a \gdr{2} radial velocity.}

As done in the \oum\ study \citep{bailer-jones2018}, we use a parallax zeropoint of $-0.029$\,mas, i.e.\ this is subtracted from all parallaxes. The true parallax zeropoint may vary and may be slightly more negative for stars brighter than $\gmag\,\simeq\,16$\,mag \cite[e.g.,][]{2019ApJ...878..136Z}, perhaps about $-0.050$\,mas. This difference is generally smaller than the parallax uncertainties for our stars. Moreover, the uncertainties in the radial velocities tend to dominate the uncertainties in our inferred encounter parameters, so the exact choice of zeropoint is not critical.

We use the same Galactic model for the orbital integration as we did in the \oum\ study.
This is a smooth, three-component axisymmetric Galaxy model. Further details can be found in \cite{bailer-jones2015}. We do not include discrete components such as molecular clouds, not least because these evolve in unknown ways on the timescales of our integration (Myr), so they cannot be correctly represented. Yet because of the limited distance horizon of our data, we generally trace orbits only for a few Myr (a few tens of pc), so the trajectories are virtually linear (demonstrated below). The exact choice of Galactic potential is therefore not very important. The accuracy of the resulting encounters is limited mostly by the stellar data.

Our method of finding encounters is identical to the one we used in \cite{bailer-jones2018} to find possible parent stars of \oum\ (see reference for more details). As in that work, in order to quantify the uncertainties in the encounter parameters, we generate 2000 surrogates for each star from its 6D covariance matrix (3D position, 3D velocity), and likewise for \borisov, and integrate all of these back in time. 

\begin{sidewaystable*}
\caption{The 14 stars that encounter \borisov\ in the past with a median encounter distance (\dencmed) 
  below 1\,pc, sorted by this value. Column 1 reports a common name (Ross, Wolf, GJ, HD, HIP, in that order) from Simbad, if available (\gdr{2} 4828141619844398464 has no entry in Simbad). Columns 3, 6, and 9 are the median encounter time \tencmed, median encounter distance \dencmed, and median encounter velocity \vencmed, respectively. The columns labeled 5\% and 95\% are the bounds of the corresponding 90\% confidence intervals. Columns 12--17 list the parallax (\parallax; corrected for the zeropoint), total proper motion ($\mu$), and radial velocity ($v_r$) along with their 1-$\sigma$ uncertainties. Columns 18, 19, and 20 are the apparent and absolute Gaia G-band magnitude (assuming zero extinction) and Gaia color respectively.
As discussed in section~\ref{sec:GJ4384}, GJ 4384 is probably in a long-period binary orbit, so the encounter parameters shown here (which neglect this) are incorrect.
$^{\sharp}$ = HD 224635? (see text), $^{\dag}$ = G 7-34, $^{\ddag}$ = van Maanen's star, $^\P$ = V* EV Lac.
\label{tab:periStats}
}
\tabcolsep=0.13cm
\begin{tabular}{*{20}{r}}
\hline
name & \gdr{2} source ID & \multicolumn{3}{c}{\tenc [kyr]} & \multicolumn{3}{c}{\denc [pc]} &  \multicolumn{3}{c}{\venc [\kms]} & \parallax & $\sigma(\parallax)$ & \propm & $\sigma(\propm)$ & \vr & $\sigma(\vr)$ & \gmag & \mg & \bprp \\
& & med & 5\% & 95\% & med & 5\% & 95\% & med & 5\% & 95\% & \multicolumn{2}{c}{mas} & \multicolumn{2}{c}{\maspyr} &  \multicolumn{2}{c}{\kms} & mag & mag & mag \\
\hline
 \object{Ross 573} & 5162123155863791744 &     -909 &     -937 &     -882 &    0.068 &    0.053 &    0.091 &    22.6 &    22.0 &    23.3 &    47.57 &     0.03 &   309.22 &     0.09 &    13.1 &     0.4 &  9.33 &  7.72 &  1.80 \myeol 
 \object{GJ 4384}$^{\sharp}$ & 2875096978193873024 &    -1523 &    -1550 &    -1496 &    0.250 &    0.218 &    0.286 &    19.1 &    18.7 &    19.4 &    33.74 &     0.04 &   126.27 &     0.06 &    -7.9 &     0.2 &  6.32 &  3.96 &  0.69 \myeol 
 \object{GJ 3270}$^{\dag}$  &  3299381442858615936 &     -444 &     -559 &     -365 &    0.459 &    0.377 &    0.580 &    32.1 &    25.5 &    39.0 &    68.58 &     0.08 &   400.33 &     0.17 &    14.0 &     4.0 & 12.25 & 11.43 &  3.13 \myeol 
 \object{HD 44867} & 3368960531532018816 &    -1424 &    -1450 &    -1396 &    0.572 &    0.360 &    1.128 &    85.9 &    85.6 &    86.2 &     8.00 &     0.09 &    48.61 &     0.11 &    71.2 &     0.2 &  6.04 &  0.56 &  1.20 \myeol 
 &  3338543951096093696 &    -2876 &    -2930 &    -2827 &    0.588 &    0.162 &    1.244 &    36.9 &    36.3 &    37.5 &     9.21 &     0.04 &    55.30 &     0.08 &    21.4 &     0.4 & 10.46 &  5.28 &  0.99 \myeol 
 \object{Wolf 28}$^{\ddag}$ &  2552928187080872832 &      -14.7 &      -15.1 &      -14.3 &    0.647 &    0.629 &    0.668 &   284.0 &   275.4 &   292.3 &   231.77 &     0.04 &  2978.29 &     0.10 &   263.0 &     4.9 & 12.31 & 14.14 &  0.61 \myeol 
 &  4828141619844398464 &    -2401 &    -2507 &    -2301 &    0.675 &    0.559 &    0.824 &    64.5 &    61.7 &    67.2 &     6.35 &     0.02 &    39.52 &     0.04 &    77.2 &     1.7 & 12.99 &  7.00 &  1.37 \myeol  
  \object{GJ 1103A} &  3085716990368639744 &      -95 &     -105 &      -86 &    0.698 &    0.634 &    0.778 &    95.7 &    86.0 &   105.0 &   107.80 &     0.09 &   801.48 &     0.17 &    93.8 &     5.7 & 11.66 & 11.82 &  3.12 \myeol 
 \object{GJ 2070} &  3073508528645520000 &     -402 &     -579 &     -307 &    0.809 &    0.615 &    1.172 &    32.5 &    22.5 &    42.6 &    74.67 &     0.07 &   479.02 &     0.12 &    34.9 &     6.2 & 11.51 & 10.87 &  2.60 \myeol 
 \object{GJ 873}$^\P$ &  1934263333784036736 &     -177 &     -178 &     -176 &    0.818 &    0.813 &    0.823 &    27.6 &    27.4 &    27.7 &   198.04 &     0.04 &   842.10 &     0.09 &     0.4 &     0.1 &  9.00 & 10.48 &  2.73 \myeol 
 &  6743117625594496512 &    -1750 &    -1783 &    -1718 &    0.835 &    0.496 &    1.220 &    69.8 &    68.6 &    70.9 &     8.00 &     0.04 &    43.84 &     0.10 &    89.6 &     0.7 & 10.41 &  4.93 &  0.95 \myeol  
 \object{HD 277756} &   207166446152692736 &    -1283 &    -1297 &    -1269 &    0.859 &    0.799 &    0.924 &    35.3 &    35.0 &    35.7 &    21.58 &     0.06 &    81.70 &     0.10 &     8.3 &     0.2 &  9.05 &  5.72 &  0.99 \myeol  
 \object{HIP 34498} &   978301126629450368 &    -5283 &    -5354 &    -5211 &    0.944 &    0.487 &    1.770 &    25.1 &    24.8 &    25.3 &     7.40 &     0.04 &    33.68 &     0.10 &     1.0 &     0.1 &  8.97 &  3.32 &  0.96 \myeol  
 \object{HD 22781} &   217334764042444288 &     -889 &     -900 &     -878 &    0.969 &    0.900 &    1.034 &    35.8 &    35.4 &    36.2 &    30.67 &     0.11 &   102.62 &     0.18 &     8.3 &     0.2 &  8.52 &  5.96 &  1.08 \myeol  
\hline
\end{tabular}
\end{sidewaystable*}

We find 14 stars that approached within 1\,pc of \borisov\ in the past.  Table~\ref{tab:periStats} shows the encounter parameters and other data on the stars. All of these have UWE of less than a few and so they are formally good astrometric solutions (if they are not wide binary systems; see section~\ref{sec:GJ4384}). We are interested in encounters that are both very close and very slow. As shown in \cite{bailer-jones2018}, a comet or asteroid ejected from a stellar system in a three-body interaction would have a characteristic ejection velocity of a few \kms\ if the third body (the stellar companion) is a giant planet. We see in Table~\ref{tab:periStats} much larger velocities, of 20\,\kms\ or more, which is only generally achievable with a stellar-mass companion.

\subsection{Ross 573}

\begin{figure*}
\begin{center}
\includegraphics[width=\textwidth, angle=0]{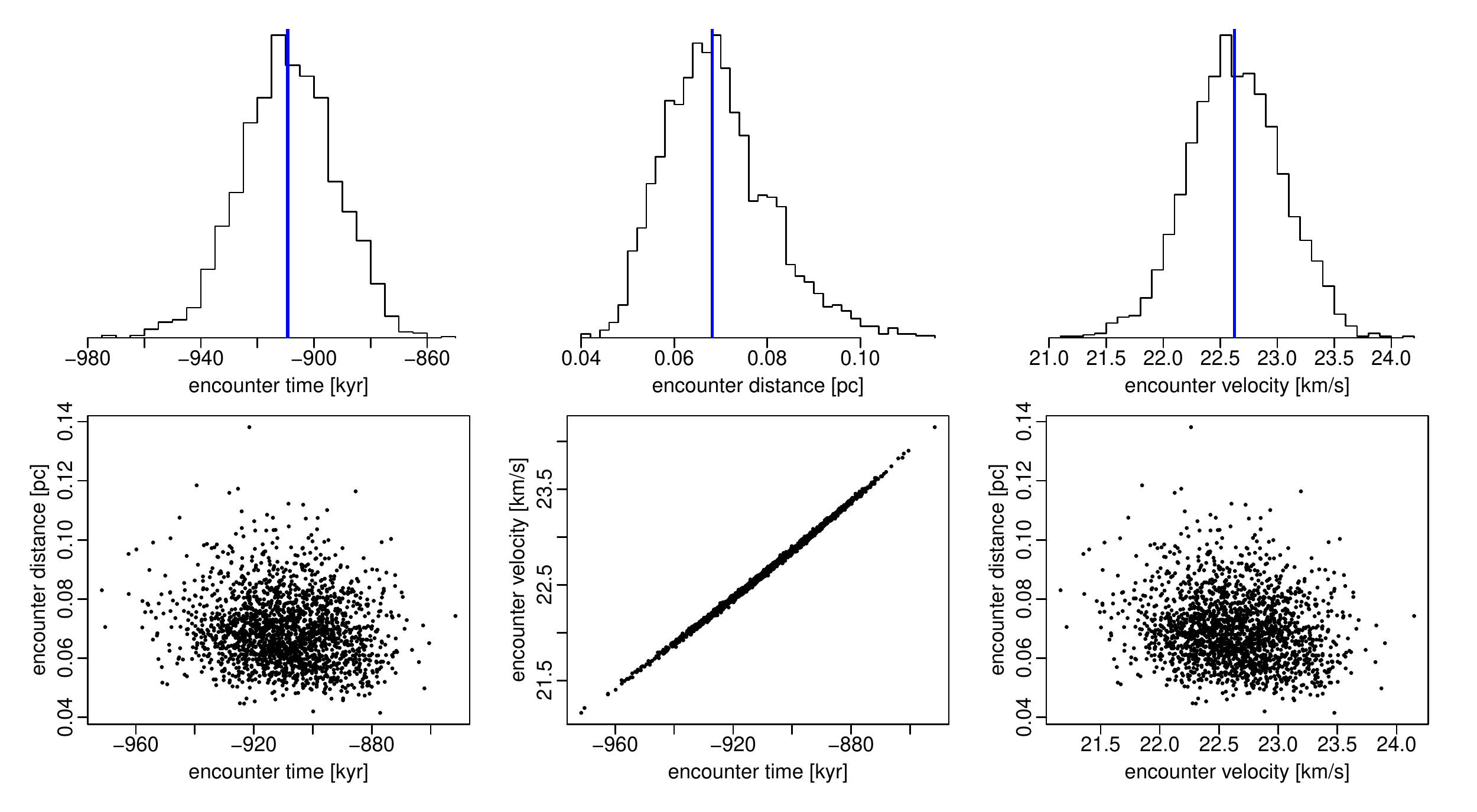}
\caption{Distribution over encounter parameters (for the 2000 surrogates) of \borisov\ asymptote solution \solused\ with \object{\gdr{2} 5162123155863791744} (= \object{Ross 573}). The vertical blue lines in the upper panels show the medians of the distributions.
\label{fig:5162123155863791744_perisamp}
}
\end{center}
\end{figure*}

The closest encounter we find passes very close, at a separation of just 0.068\,pc (around 14\,000\,au) with a 90\% confidence interval (CI) of 0.053--0.091\,pc. The encounter took place 910\,kyr ago at a relative speed of 23\,\kms. Figure~\ref{fig:5162123155863791744_perisamp} shows the distribution of the surrogates. This object, \object{\gdr{2} 5162123155863791744}, has various other names, including \object{Ross 573}, and it is listed in Simbad as being an M0 dwarf with an apparent V-band magnitude of 10.0\,mag. These characteristics agree with its \gdr{2}\ apparent and absolute magnitudes and color (listed in Table~\ref{tab:periStats}), as well as its effective temperature (\teff) of 4050\,K (68\% CI 3980--4180\,K). This is a high proper-motion star (310\,\maspyr) that is currently 21\,pc from the Sun. None of the \gdr{2} quality metrics suggest the \gaia\ astrometry is problematic (e.g.\ UWE\,=\,1.3). No nearby companions are listed in Simbad and none are apparent in images in Aladin \citep{2000A&AS..143...33B}, so there is no evidence that the astrometric solution is compromised by duplicity. There are seven good HARPS spectra in the ESO archive taken over a period of nine years. Their radial velocities are consistent within 20\,\ms, which is further evidence for a lack of significant binarity. The HARPS radial velocity is $14.1105 \pm 0.0068$\,\kms. If we use this in the orbital integration, the encounter parameters hardly change: \dencmed\ decreases by 0.0039\,pc and \tencmed\ increases by 41\,kyr.

Assuming \object{Ross 573} has an Oort cloud and that the radius of such an Oort cloud is limited by the Galactic tide, then the maximum radius scales with the size of the star's Hill sphere in the Galactic potential \citep{2018MNRAS.473.5432H}.  As \object{Ross 573} is very close to the Sun, the scaling factor is simply $(M_{\rm {star}}/\msun)^{1/3}$. With $M_{\rm {star}} = 0.7$\,\msun\ from the spectral type, this scaling factor is 0.9. Thus the maximum radius for \object{Ross 573}'s Oort cloud is similar to that of the Sun, which is about $10^5$\,au. The encounter with \borisov\ is much less than this and so, from this point of view, \object{Ross 573} is a plausible origin for this exo-comet.

Using the median encounter parameters and the adopted mass of \object{Ross 573}, we compute that \borisov\ was deflected by 36\arcsec\ by this encounter.  The typical relative velocities of, and timescales to, other close encounters in Table~\ref{tab:periStats} are 20\,\kms\ and 500\,kyr respectively, during which \borisov\ will have travelled 10\,pc. Over this distance the deflection corresponds to a lateral displacement of 0.002\,pc. That is, our computation of other close encounters could be wrong by up to this amount due to our neglect of two-body effects. This is much smaller than the precision of the inferred encounter separations (and the other encounters induce even smaller errors).

The trajectory of \object{Ross 573}, and most of the other close encounters in Table~\ref{tab:periStats}, is so short that the exact Galactic potential has little impact on the encounter parameters. As an extreme test we set the potential to zero (i.e., using linear motions) and recompute the encounter parameters, again using all the surrogates. The median encounter
time, distance, and velocity then change by $-0.9$\,kyr, $-0.0036$\,pc, and $-0.03$\,\kms, respectively. These are much smaller than the uncertainties.

\object{Ross 573} was also identified as a close encounter by \cite{hallatt2019} who infer a similar time and speed, but with a considerably larger encounter distance of 0.64\,pc (90\% CI of 0.60--0.69\,pc). Their study used a combination of sources for the data on the stars, and it is not stated exactly which data they used for each object, although it seems likely that they used \gdr{2} here. They use a slightly different Galaxy model, but given that our encounter parameters are hardly changed when we just assume linear motions,
this is unlikely to be the reason for the discrepancy. A more likely explanation is the use of a different asymptotic trajectory for \borisov, in particular the velocity amplitude (see Section~\ref{sec:trajectory}). \cite{hallatt2019} used an early gravity-only solution obtained from JPL on 2 October 2019. Ours is based on a much longer data arc and accounts for the action of non-gravitational forces, so it should be more accurate.

\subsection{GJ 4384}\label{sec:GJ4384}

Our second closest encounter is \object{GJ 4384}. To be more precise, the result in Table~\ref{tab:periStats} uses astrometry for \object{\gdr{2} 2875096978193873024} and the Simbad-listed radial velocity of $-7.9 \pm 0.2$\,\kms (there is no \gdr{2} radial velocity), as these objects match to within 0.1\arcsec. Simbad lists a second source, \object{HD 224635}, as being only 0.01\arcsec\ from \object{GJ 4384}. However, from the data on these sources from various catalogs and publications listed on Simbad, we conclude that these two identifiers actually refer to the same single source. \cite{2012A&A...546A..69M}, for example, in a study of binary systems, does not flag this as being either an eclipsing or spectroscopic binary star. Also, it is not identified as an unresolved binary in Hipparcos \citep{1997ESASP1200.....P}. We equate this star with \object{ADS 17149A}, \object{HIP 118281}, and \object{WDS J23595+3343A}, and refer to it here as s024 for short. If, instead, we use the radial velocity listed in Simbad for \object{HD 224635}, which is $-7.7 \pm 2.0$\,\kms, then the encounter parameters shown in Table~\ref{tab:periStats} are hardly changed (\dencmed\ by 0.0008\,pc and \tencmed\ by 17\,kyr).

\object{GJ 4384} does, however, have a companion 2.475\arcsec\ away, \object{Gaia DR2 2875096978193654528} (s528 for short), which also goes by the names \object{HD 224636}, \object{ADS 17149B}, and \object{WDS J23595+3343B}.

These two stars, s024 and s528, both have good astrometric quality indicators, so their proximity should not have corrupted their astrometry (in agreement with expectations about \gaia\ for this separation).
But they do have very similar parallaxes (the average is 33.82\,mas and the difference 0.04\,mas)
and radial velocities (Simbad gives $-4.6 \pm 2.0$\,\kms\ for \object{HD 224636}), so they are almost certainly a physical binary.  If they were in a common plane perpendicular to the line-of-sight, their physical separation would be 73\,au and their relative velocity 4.8\,\kms. This compares to a relative orbital velocity of two solar-mass stars in a mutual circular orbit of this radius of 4.9\,\kms (and their period would be 440\,yr).  As their proper motion difference of 35\,\maspyr\ is about a quarter of their absolute proper motions, the velocity of their center-of-mass could differ significantly from that of either component.

If we knew the mases of both components of the binary, we could solve for both the center-of-mass and the Keplerian orbital elements (6+6 parameters from the 3D position and 3D velocity for each star).  The spectral type of this pair is given by \cite{1955ApJ...121..337S} as F8+G1, with F8 presumably referring to \object{GJ 4384} because this is slightly brighter and bluer (by 0.3\,mag and 0.05\,mag respectively in \gdr{2}).
This agrees with their \teff\ and positions in the color-absolute magnitude diagram, as given by \gdr{2}.  We adopt a mass of 1\,\msun\ for each component to compute the position and velocity of the center-of-mass.  Integrating the center-of-mass's motion back in time, we find the time, distance, and relative speed of its closest encounter with \borisov\ to be -1380\,kyr, 3.64\,pc, and 20.9\,\kms\ respectively.  This is considerably more distant than the encounter found when neglecting the binary motion (the second line of Table~\ref{tab:periStats}).
We do not compute the covariance on these parameters due to an inadequately characterized correlation in the \gaia\ astrometric uncertainties for sources close to one another. The largest uncertainty is the radial velocity of s528. If we change this by $\pm 2\sigma$, the encounter distance remains between 3.3 and 4.0\,pc. We conclude that the binary motion renders the \object{GJ 4384}--\object\object{HD 224636} system a much more distant encounter than is shown in Table~\ref{tab:periStats}. Although a binary system is an attractive option for ejecting \borisov, this encounter distance is far larger than the size of the system.

We note that \object{GJ 4384} is listed as the second closest encounter to \borisov\ as found by \cite{hallatt2019} with similar encounter parameters to ours (\dencmed\,=\,0.32\,pc) when we neglect binarity.

\subsection{Other candidates}

We note that several of our other close encounters are binary systems. One of these, \object{HD 22781}, has a planetary companion \citep{2012A&A...538A.113D}, but the encounter velocity of 36\,\kms\ is rather high for this planet to be plausibly involved in ejecting \borisov.

It is worth examining in addition those encounters beyond 1\,pc (not listed in Table~\ref{tab:periStats}) that are particularly slow. The slowest encounter we find within 5\,pc is with \object{\gdr{2} 4769698316220254592} (=\,\object{HD 34327}), at 3.9\,\kms (90\% CI 3.7--4.2\,\kms) but at 4.8\,pc, 14.7\,Myr ago.
The next slowest encounter is nominally \object{\gdr{2} 2007876324466455424} (=\,\object{HD 239960A}) with \vencmed\,=\,5.6\,\kms\ (and \denc\ between 1 and 3.5\,pc). However, because of the large relative uncertainty in its velocity relative to \borisov, mostly arising from the radial velocity uncertainty, this encounter solution is poorly constrained.

The closest encounter that \cite{hallatt2019} find to \borisov\ is with \object{GJ 3270}, at \dencmed\,=\,0.21\,pc (90\% CI 0.08--0.32\,pc) and \vencmed\,=\,33\,\kms. Our solution puts this encounter further away, with a 90\% probability of approaching between 0.377 and 0.580\,pc (see Table~\ref{tab:periStats}), and only a 1\% probability that its approach is closer than 0.35\,pc. The third closest encounter from \cite{hallatt2019}, at 0.32--0.55\,pc (90\% CI), is with \object{2MASS J03552337+1133437} (=\,\object{Gaia DR2 3303349202364648320}). Our solutions put this much further away, at 1.07\,pc (90\% CI 0.98--1.17\,pc), so it is not listed in our table.

\section{Discussion}
\label{sec:discussion}

The proximity of \object{Ross 573}'s encounter with \borisov\ does not, by itself, indicate that the former is the origin of the latter. To put this proximity in context, we can estimate how often a star would happen to pass this closely to \borisov. Using \gdr{2}, \cite{2018A&A...616A..37B} computed that the rate at which stars approach within 1\,pc of the Sun is $19.7\pm 2.2$ per Myr. This rate includes all types of stars and is corrected for \gaia's incompleteness. \cite{2018A&A...609A...8B} show that this rate scales quadratically with the encounter distance. To first order, this same rate should apply for any object in the solar neighborhood, including \borisov. Thus, the rate of encounters of stars within 0.067\,pc of \borisov\ is expected to be $0.088\pm 0.010$ per Myr, which is one every 11.3\,Myr. This is 12 times longer than the time back to \borisov's encounter with \object{Ross 573}, suggesting this was unlikely to be just a chance encounter. However, because we only consider \gdr{2} sources in the present study, a more appropriate comparison is with the encounter rate just for stars in \gdr{2}, that is, before correcting for \gaia's incompleteness. This is 0.0083 encounters per Myr within 0.067\,pc
\citep{2018A&A...616A..37B}, which corresponds to an expected time of 120\,Myr lapsing before such a close encounter, thus making it even less likely this was just a chance encounter.

An alternative estimate of the probability of coincidence can be made using Poisson statistics. Assuming a space density of stars (of potential origin) of 0.1 per cubic parsec,
then the average number of flybys within 0.067\,pc of \borisov\ over the distance it has has travelled from \object{Ross 573} to the Sun (30\,pc) is $\lambda=0.044$. The probability of \borisov\ having experienced an unassociated flyby as close as this is, therefore, $1-e^{-\lambda}=0.04$.

In their study of \oum, \cite{bailer-jones2018} show that a comet can be ejected from a stellar system containing a stellar or giant planet companion with velocities of a few km/s.  For the solar system -- that is, a system dominated by a 1\,\msun\ star and a 1\,\mjup\ planet at 5\,au -- they find that the maximum ejection velocity of the comet was 17.4\,\kms, which is not enough for \borisov.  A more massive companion can eject at higher velocities, up to many tens of \kms\ for masses up to the hydrogen-burning limit. Yet to achieve the required ejection velocity of 23\,\kms\ from \object{Ross 573}, a close interaction would be required, which is, a priori, less likely. Furthermore, there is no indication from the \gdr{2} astrometry, HARPS spectroscopy, or the literature that \object{Ross 573} is a binary. An alternative ejection mechanism is the close passage of another (unbounded) star. We searched \gdr{2} for this. The only star coming within 1\,pc of \object{Ross 573} in a time window around the \borisov--\object{Ross 573} encounter is \object{\gdr{2} 347051160157809920}, with encounter parameters \tenc\,=\,$-$1080\,kyr, \denc\,=\,0.72\,pc, \venc\,=\,41\,\kms, which is neither particularly close nor slow.

Our investigation is based on the assumption that \borisov\ is not older than about 10\,Myr. This is because \borisov\ travels 33\,pc per Myr relative to the Sun, yet the \gdr{2}+Simbad radial velocity sample becomes severely incomplete after a few hundred parsec. Currently, there is no age established for \borisov. Although its relatively large speed with respect to the local standard of rest -- its pre-encounter space velocity is $(U,V,W)=(33.1, -6.8, 8.3)$\,\kms\ in our Galaxy model
-- could suggest it is old, this is only a weak indication. Future \gaia\ data releases will not extend the reliable traceback horizon substantially because of the relatively bright magnitude limit on its radial velocities.  Even if we had a much deeper six-parameter stellar catalogue, limitations remain on how far back we can trace orbits, due to inaccuracies in our knowledge of the Galactic potential and its evolution \citep[e.g.,][]{bailer-jones2018, 2018ApJ...852L..13Z, hallatt2019}.

In this study, we ask whether there are close encounters to specific stars among the available data. We have not attempted
to estimate the overall rate of close encounters to \borisov\ by correcting for this incompleteness \cite[as done for the Sun, for example, by][]{2001A&A...379..634G,2018A&A...609A...8B, 2018A&A...616A..37B}; nor have we corrected for binary motion. Even though this could be done in a few cases, the vast majority of binaries are either unidentified as binaries or could not be corrected using the available data.

\section{Conclusions}
\label{sec:conclusions}

We have found a very close (14\,000\,au; 90\% CI 11\,000--18\,600\,au) encounter between \borisov\ and the M0V star \object{Ross 573}  that took place around 900\,kyr ago. If \borisov\ was ejected from orbit around this star via a three-body interaction, then assuming it has undergone no other encounter since, it must have been ejected at around 23\,\kms. This is rather fast for ejection by a massive companion, for which there is also no observational evidence. We have not found a good candidate for a second star passing close to \object{Ross 573} that could potentially have ejected \borisov. Thus although the probability of such a close encounter over the distance \borisov\ has traveled from this star to the Sun is low (0.04), there is no obvious mechanism to eject \borisov\ from the host star.
A second close approach that initially seemed interesting took place with the F8V star \object{GJ 4384}, 1500\,kyr ago at 0.25\,pc and 19\,\kms.
However, when taking into account its orbital motion with its solar-mass companion at a projected separation of 73\,au, the encounter separation to the center-of-mass of this system increases to over 3\,pc.

\begin{acknowledgements}
We thank Olivier Hainaut for searching the ESO archive for data on \object{Ross 573}. This research has made use of:  data and services provided by the International Astronomical Union's Minor Planet Center; the Simbad and Aladin services provided CDS, Strasbourg;  data from the European Space Agency (ESA) mission \gaia\ (\url{http://www.cosmos.esa.int/gaia}), processed by the \gaia\ Data Processing and Analysis Consortium (DPAC, \url{http://www.cosmos.esa.int/web/gaia/dpac}). Funding for the DPAC has been provided by national institutions, in particular the institutions participating in the \gaia\ Multilateral Agreement.
DF conducted this research at the Jet Propulsion Laboratory, California Institute of Technology, under a contract with NASA. 
KJM acknowledges support through awards from NASA 80NSSC18K0853. 
\end{acknowledgements}

\end{CJK*}

\bibliographystyle{aa}
\bibliography{2I}

\begin{thebibliography}{35}
\expandafter\ifx\csname natexlab\endcsname\relax\def\natexlab#1{#1}\fi

\bibitem[{{Bailer-Jones}(2015)}]{bailer-jones2015}
{Bailer-Jones}, C.~A.~L. 2015, \aap, 575, A35

\bibitem[{{Bailer-Jones}(2018)}]{2018A&A...609A...8B}
{Bailer-Jones}, C.~A.~L. 2018, \aap, 609, A8

\bibitem[{{Bailer-Jones} {et~al.}(2018{\natexlab{a}}){Bailer-Jones},
  {Farnocchia}, {Meech}, {Brasser}, {Micheli}, {Chakrabarti}, {Buie}, \&
  {Hainaut}}]{bailer-jones2018}
{Bailer-Jones}, C. A.~L., {Farnocchia}, D., {Meech}, K.~J., {et~al.}
  2018{\natexlab{a}}, \aj, 156, 205

\bibitem[{{Bailer-Jones} {et~al.}(2018{\natexlab{b}}){Bailer-Jones}, {Rybizki},
  {Andrae}, \& {Fouesneau}}]{2018A&A...616A..37B}
{Bailer-Jones}, C.~A.~L., {Rybizki}, J., {Andrae}, R., \& {Fouesneau}, M.
  2018{\natexlab{b}}, \aap, 616, A37

\bibitem[{{Belton} {et~al.}(2018){Belton}, {Hainaut}, {Meech}, {Mueller},
  {Kleyna}, {Weaver}, {Buie}, {Drahus}, {Guzik}, {Wainscoat}, {Waniak}, {Hand
  zlik}, {Kurowski}, {Xu}, {Sheppard}, {Micheli}, {Ebeling}, \&
  {Keane}}]{belton2018}
{Belton}, M. J.~S., {Hainaut}, O.~R., {Meech}, K.~J., {et~al.} 2018, \apjl,
  856, L21

\bibitem[{{Bonnarel} {et~al.}(2000){Bonnarel}, {Fernique}, {Bienaym{\'e}},
  {Egret}, {Genova}, {Louys}, {Ochsenbein}, {Wenger}, \&
  {Bartlett}}]{2000A&AS..143...33B}
{Bonnarel}, F., {Fernique}, P., {Bienaym{\'e}}, O., {et~al.} 2000, \aaps, 143,
  33

\bibitem[{{CBET}(2019)}]{borisov2019}
{CBET}. 2019, Central Bureau for Astronomical Telegrams, CBET4666

\bibitem[{{Chesley} \& {Yeomans}(2005)}]{chesley2005}
{Chesley}, S.~R. \& {Yeomans}, D.~K. 2005, in IAU Colloq. 197: Dynamics of
  Populations of Planetary Systems, ed. Z.~{Kne{\v{z}}evi{\'c}} \& A.~{Milani},
  289--302

\bibitem[{{D{\'\i}az} {et~al.}(2012){D{\'\i}az}, {Santerne}, {Sahlmann},
  {H{\'e}brard}, {Eggenberger}, {Santos}, {Moutou}, {Arnold}, {Boisse},
  {Bonfils}, {Bouchy}, {Delfosse}, {Desort}, {Ehrenreich}, {Forveille},
  {Lagrange}, {Lovis}, {Pepe}, {Perrier}, {Queloz}, {S{\'e}gransan}, {Udry}, \&
  {Vidal-Madjar}}]{2012A&A...538A.113D}
{D{\'\i}az}, R.~F., {Santerne}, A., {Sahlmann}, J., {et~al.} 2012, \aap, 538,
  A113

\bibitem[{{Farnocchia} {et~al.}(2015){Farnocchia}, {Chesley}, {Milani},
  {Gronchi}, \& {Chodas}}]{2015aste.book..815F}
{Farnocchia}, D., {Chesley}, S.~R., {Milani}, A., {Gronchi}, G.~F., \&
  {Chodas}, P.~W. 2015, in Asteroids IV, ed. P.~{Michel}, F.~{DeMeo}, \&
  W.~{Bottke} (Tucson: University of Arizona Press), 815--834

\bibitem[{{Fitzsimmons} {et~al.}(2019){Fitzsimmons}, {Hainaut}, {Meech},
  {Jehin}, {Moulane}, {Opitom}, {Yang}, {Keane}, {Kleyna}, {Micheli}, \&
  {Snodgrass}}]{fitzsimmons2019}
{Fitzsimmons}, A., {Hainaut}, O., {Meech}, K.~J., {et~al.} 2019, \apjl, 885, L9

\bibitem[{{Fitzsimmons} {et~al.}(2018){Fitzsimmons}, {Snodgrass}, {Rozitis},
  {Yang}, {Hyland}, {Seccull}, {Bannister}, {Fraser}, {Jedicke}, \&
  {Lacerda}}]{fitzsimmons2018}
{Fitzsimmons}, A., {Snodgrass}, C., {Rozitis}, B., {et~al.} 2018, Nature
  Astronomy, 2, 133

\bibitem[{{Gaia Collaboration}(2018)}]{2018A&A...616A...1G}
{Gaia Collaboration}. 2018, \aap, 616, A1

\bibitem[{{Garc{\'{\i}}a-S{\'a}nchez}
  {et~al.}(2001){Garc{\'{\i}}a-S{\'a}nchez}, {Weissman}, {Preston}, {Jones},
  {Lestrade}, {Latham}, {Stefanik}, \& {Paredes}}]{2001A&A...379..634G}
{Garc{\'{\i}}a-S{\'a}nchez}, J., {Weissman}, P.~R., {Preston}, R.~A., {et~al.}
  2001, \aap, 379, 634

\bibitem[{{Hallatt} \& {Wiegert}(2019)}]{hallatt2019}
{Hallatt}, T. \& {Wiegert}, P. 2019, arXiv e-prints, arXiv:1911.02473v1

\bibitem[{{Hanse} {et~al.}(2018){Hanse}, {J{\'\i}lkov{\'a}}, {Portegies Zwart},
  \& {Pelupessy}}]{2018MNRAS.473.5432H}
{Hanse}, J., {J{\'\i}lkov{\'a}}, L., {Portegies Zwart}, S.~F., \& {Pelupessy},
  F.~I. 2018, \mnras, 473, 5432

\bibitem[{{Jewitt} \& {Luu}(2019)}]{2019ApJ...886L..29J}
{Jewitt}, D. \& {Luu}, J. 2019, \apjl, 886, L29

\bibitem[{{Malkov} {et~al.}(2012){Malkov}, {Tamazian}, {Docobo}, \&
  {Chulkov}}]{2012A&A...546A..69M}
{Malkov}, O.~Y., {Tamazian}, V.~S., {Docobo}, J.~A., \& {Chulkov}, D.~A. 2012,
  \aap, 546, A69

\bibitem[{{Marsden} {et~al.}(1973){Marsden}, {Sekanina}, \&
  {Yeomans}}]{marsden1973}
{Marsden}, B.~G., {Sekanina}, Z., \& {Yeomans}, D.~K. 1973, \aj, 78, 211

\bibitem[{{Meech} \& {Svoren}(2004)}]{2004come.book..317M}
{Meech}, K.~J. \& {Svoren}, J. 2004, in Comets II, ed. M.~C. {Festou}, H.~U.
  {Keller}, \& H.~A. {Weaver} (Tucson: University of Arizona Press), 317

\bibitem[{{Meech} {et~al.}(2017){Meech}, {Weryk}, {Micheli}, {Kleyna},
  {Hainaut}, {Jedicke}, {Wainscoat}, {Chambers}, {Keane}, {Petric}, {Denneau},
  {Magnier}, {Berger}, {Huber}, {Flewelling}, {Waters}, {Schunova-Lilly}, \&
  {Chastel}}]{meech2017}
{Meech}, K.~J., {Weryk}, R., {Micheli}, M., {et~al.} 2017, \nat, 552, 378

\bibitem[{{Micheli} {et~al.}(2018){Micheli}, {Farnocchia}, {Meech}, {Buie},
  {Hainaut}, {Prialnik}, {Sch{\"o}rghofer}, {Weaver}, {Chodas}, {Kleyna},
  {Weryk}, {Wainscoat}, {Ebeling}, {Keane}, {Chambers}, {Koschny}, \&
  {Petropoulos}}]{micheli2018}
{Micheli}, M., {Farnocchia}, D., {Meech}, K.~J., {et~al.} 2018, \nat, 559, 223

\bibitem[{{Minor Planet Center}(2019{\natexlab{a}})}]{mpcR106}
{Minor Planet Center}. 2019{\natexlab{a}}, Minor Planet Electronic Circulars,
  2019-R106

\bibitem[{{Minor Planet Center}(2019{\natexlab{b}})}]{mpcS71}
{Minor Planet Center}. 2019{\natexlab{b}}, Minor Planet Electronic Circulars,
  2019-S71

\bibitem[{{Opitom} {et~al.}(2019){Opitom}, {Fitzsimmons}, {Jehin}, {Moulane},
  {Hainaut}, {Meech}, {Yang}, {Snodgrass}, {Micheli}, {Keane}, {Benkhaldoun},
  \& {Kleyna}}]{Opitom2019}
{Opitom}, C., {Fitzsimmons}, A., {Jehin}, E., {et~al.} 2019, \aap, 631, L8

\bibitem[{{Perryman} \& {ESA}(1997)}]{1997ESASP1200.....P}
{Perryman}, M.~A.~C. \& {ESA}, eds. 1997, ESA Special Publication, Vol. 1200,
  {The HIPPARCOS and TYCHO catalogues. Astrometric and photometric star
  catalogues derived from the ESA HIPPARCOS Space Astrometry Mission}

\bibitem[{{Struve} \& {Franklin}(1955)}]{1955ApJ...121..337S}
{Struve}, O. \& {Franklin}, K.~L. 1955, \apj, 121, 337

\bibitem[{{Wenger} {et~al.}(2000){Wenger}, {Ochsenbein}, {Egret}, {Dubois},
  {Bonnarel}, {Borde}, {Genova}, {Jasniewicz}, {Lalo{\"e}}, {Lesteven}, \&
  {Monier}}]{2000A&AS..143....9W}
{Wenger}, M., {Ochsenbein}, F., {Egret}, D., {et~al.} 2000, \aaps, 143, 9

\bibitem[{{Whipple}(1950)}]{1950ApJ...111..375W}
{Whipple}, F.~L. 1950, \apj, 111, 375

\bibitem[{{Whipple}(1951)}]{1951ApJ...113..464W}
{Whipple}, F.~L. 1951, \apj, 113, 464

\bibitem[{{Ye} {et~al.}(2019){Ye}, {Kelley}, {Bolin}, {Bodewits}, {Farnocchia},
  {Masci}, {Meech}, {Micheli}, {Weryk}, {Bellm}, {Christensen}, {Dekany},
  {Delacroix}, {Graham}, {Kulkarni}, {Laher}, {Rusholme}, \& {Smith}}]{ye2019}
{Ye}, Q., {Kelley}, M. S.~P., {Bolin}, B.~T., {et~al.} 2019, arXiv e-prints,
  arXiv:1911.05902

\bibitem[{{Ye} {et~al.}(2017){Ye}, {Zhang}, {Kelley}, \& {Brown}}]{ye2017}
{Ye}, Q.-Z., {Zhang}, Q., {Kelley}, M. S.~P., \& {Brown}, P.~G. 2017, \apjl,
  851, L5

\bibitem[{{Yeomans} \& {Chodas}(1989)}]{1989AJ.....98.1083Y}
{Yeomans}, D.~K. \& {Chodas}, P.~W. 1989, \aj, 98, 1083

\bibitem[{{Zhang}(2018)}]{2018ApJ...852L..13Z}
{Zhang}, Q. 2018, \apjl, 852, L13

\bibitem[{{Zinn} {et~al.}(2019){Zinn}, {Pinsonneault}, {Huber}, \&
  {Stello}}]{2019ApJ...878..136Z}
{Zinn}, J.~C., {Pinsonneault}, M.~H., {Huber}, D., \& {Stello}, D. 2019, \apj,
  878, 136

\end{thebibliography}

\end{document}